\documentclass[twocolumn,secnumarabic,amssymb, amsmath, nofootinbib,tightenlines,
nobibnotes, aps, prl,epsfig]{revtex4}
\usepackage{graphicx}
\usepackage{dcolumn}
\usepackage{bm}
\begin{document}
\preprint{APS/123-QED}
\title{The exponent of the longitudinal structure function $F_{L}$ at low $x$}

\author{G.R.Boroun}%
 \email{grboroun@gmail.com; boroun@razi.ac.ir }
\affiliation{ Physics Department, Razi University, Kermanshah
67149, Iran}
\date{\today}
\begin{abstract}

We present a set of formula to extract exponents of the
longitudinal structure function  and reduced cross section from
the Regge-like behavior at small $x$. The exponents are found to
be independent of $Q^{2}$ at NNLO analysis. As a result, we show
that the  reduced cross section exponents do not have the same
behavior at some values of $x$. This difference predicts the non-
linear effects and some evidence for shadowing and antishadowing
at LHeC. Also the ratio $\frac{F_{2}}{\sigma}$ is calculated and
compared with the corresponding HERA data. Our calculations show a
very good agreement with the DIS experimental
data throughout the small values of $x$.\\

\end{abstract}
 \pacs{***}
\maketitle
\tableofcontents
\subsection{1. Introduction}
 The appropriate framework for the theoretical description of the
 small-$x$ behavior of the structure functions is the Regge
 approach. The Regge theory gives a good description of the structure
functions, where the high-energy scattering can be described by
power-like behavior at small-$x$. The following parameterization
of the deep inelastic scattering structure function
$F_{2}(x,Q^{2})$ defined by
\begin{eqnarray}
F_{2}(x,Q^{2})\sim \sum_{i}A_{i}(Q^{2})x^{-\lambda_{i}},
\end{eqnarray}
that the singlet part of the structure function is controlled by
pomeron exchange. Here the $i=0$ term is hard-pomeron  and $i=1$
is soft-pomeron exchange [1-2]. The effective intercept behavior,
at small values of $x$, exhibited for the fast growth of the
singlet structure function. The exponent $\lambda_{s}$ is found to
be $\simeq 0.33$ in Refs.[3-4]. It can be recast into the symbolic
form as
\begin{eqnarray}
F_{2}(x,Q^{2})=A_{s}(Q^{2})x^{-\lambda_{s}}.
\end{eqnarray}
According to the Regge theory the charm component
$F_{2}^{c}(x,Q^{2})$ of $F_{2}(x,Q^{2})$ is governed entirely by
hard pomeron exchange. In perturbative quantum chromodynamics
(pQCD) the charmed quark originates from a gluon in the proton.
Therefore the small-$x$ behavior of the gluon distribution
function is dominated with hard-pomeron intercept as
\begin{eqnarray}
G(x,Q^{2})=A_{g}(Q^{2})x^{-\lambda_{g}},
\end{eqnarray}
where   $\lambda_{g}\simeq 0.42$ [1-6]. This implies that the
gluon distribution is dominated by the hard pomeron behavior.
Indeed this steep behavior of the gluon distribution generates a
similar steep behavior of $F_{2}$ at small $x$ where
$\lambda_{s}\neq
\lambda_{g}$ in high-order corrections.\\
We now consider the proton$^{,}$s longitudinal structure function
$F_{L}(x,Q^{2})$. It is known that the dominant source for the
longitudinal structure function, at small-$x$, is the gluon
density. It is become traditional to believe that the longitudinal
intercept has the same behavior of the gluon intercept. As a
alternative, one can study the power-like behavior of $F_{L}$ via
analytical solutions of the evolution equations. It is tempting,
however, to explore the possibility of obtaining analytical
solutions of the longitudinal intercept in the restricted domain
of small $x$ at least. In this paper we suggest the power like
behavior of the longitudinal structure function as
\begin{eqnarray}
F_{L}(x,Q^{2})=A_{L}(Q^{2})x^{-\lambda_{L}},
\end{eqnarray}
at high-order corrections [7].\\
In pQCD, the Altarelli-Martinelli equation for longitudinal
structure function in terms of coefficient functions is given by
[8-9]
\begin{eqnarray}
\mathrm{gluonic~term}:~~~~~~~~~~~~~~~~~~~~~~~~~~~~~~~~~~~~~~~~~~~~~~~\nonumber\\
x^{-1}F_{L}=<e^{2}>C_{L,g}\otimes
g,\nonumber\\
\mathrm{singlet~+~ gluon~terms}:n_{f}=4~~~~~~~~~~~~~~~~~~~~~~~~~~\nonumber\\
x^{-1}F_{L}=<e^{2}>(C_{L,q}\otimes q_{s}+C_{L,g}\otimes g),\nonumber\\
\mathrm{light~ quarks~+~ heavy~terms}:n_{f}=3~~~~~~~~~~~~~~~~~\nonumber\\
x^{-1}F_{L}=<e^{2}>(C_{L,q}\otimes q_{s}+C_{L,g}\otimes
g)+x^{-1}F^{c}_{L}.
\end{eqnarray}
At small $x$, the nonsinglet contribution $F_{L}^{ns}$ is
negligible and can be ignored. Here $q_{s}$ and $g$ are the
flavour singlet and gluon distribution function, where $<e^{2}>$
stand for the average of the charge $e^{2}$ for the active quark
flavours ($<e^{2}>=n^{-1}_{f}\sum_{i=1}^{n_{f}}e_{i}^{2}$) and
$n_{f}$ denotes the number of active light flavours.\\
$C_{L,q \&
g}(\alpha_{s},x)=\sum_{n=1}(\frac{\alpha_{s}}{4\pi})^{n}C^{(n)}_{L,q
\& g}(x)$ [9] where $n$ denotes the order in $\alpha_{s}$ as at
NNLO analysis the running coupling constant has the following form
\begin{eqnarray}
\alpha_{s}(t)&=&\frac{4\pi}{\beta_{0}t}[1-\frac{\beta_{1}}{\beta_{0}^{2}}\frac{\ln{t}}{t}\nonumber\\
&&+\frac{1}{\beta_{0}^{3}t^{2}}\{\frac{\beta_{1}^{2}}{\beta_{0}}(\ln^{2}{t}-\ln{t}-1)+\beta_{2}
\} ].
\end{eqnarray}
Here $\beta^{,}s$ are the high-order corrections to the QCD
$\beta$-function and $t=\ln\frac{Q^{2}}{\Lambda^{2}}$ where
$\Lambda$ is the QCD cut-off parameter. In Eq.(5) we use the NLO
expression for the longitudinal charm structure function
$F_{L}^{c}$ [10-11] where the charm cross-section is generated by
photon-gluon fusion. This is called the fixed flavour number
scheme (FFNS) and incorporates the correct threshold behavior for
$Q^{2}\sim m_{c}^{2}$ and extended to the zero mass variable
flavour number scheme (ZM-VFNS) above this threshold [12]. In the
framework of this scheme we consider the heavy flavor physics in
the DGLAP [13] dynamics. Further simplification is obtained by
neglecting the contributions caused
by incoming light quark and antiquarks at small values of $x$.\\
The contribution of the longitudinal structure function $F_{L}$ to
the cross section can be sizeable only at large values of the
inelasticity $y$ [14]. The reduced cross section is defined as
\begin{eqnarray}
\sigma_{r}\equiv F_{2}(x,Q^{2})-\frac{y^{2}}{Y_{+}}F_{L}(x,Q^{2}),
\end{eqnarray}
where $Y_{+}=1+(1-y)^2$. At small-$y$ the relation
$\sigma_{r}\approx x^{-\lambda_{s}}$ holds to a very good
approximation as the cross section rises with decreasing $x$
($Q^{2}=sxy$ where $s$ is the center of mass energy squared).
However, at very high-$y$ a characteristic bending of the cross
section is attributed to the longitudinal
structure function contribution [15-16]. \\
In this paper, we suggest analytical solutions of the high-order
corrections for the longitudinal structure function exponent at
small $x$. The results have been included in the  reduced cross
section exponent. The behavior of these exponents are compared
with the gluon and singlet exponents where hard pomeron is dominant.\\

\subsection{2. Behavior of $F_{L}$}
\subsection{2.1. Gluonic term:}
The perturbative predictions for the gluonic longitudinal
structure function can be written as
\begin{eqnarray}
F^{g}_{L}(x,Q^{2})=<e^{2}>C_{L,g}(\alpha_{s},x)\otimes G(x,Q^{2}),
\end{eqnarray}
where $G(x,Q^{2})=xg(x,Q^{2})$. The evolution of ${\partial
F^{g}_{L}(x,Q^{2})}/{\partial \ln{x}}$ at fixed $Q^{2}$  is
obtained by the following form
\begin{eqnarray}
\frac{\partial F^{g}_{L}(x,Q^{2})}{\partial
\ln{x}}&=&<e^{2}>\{\frac{\partial G(x,Q^{2})}{\partial
\ln{x}}(C_{L,g}(x,\alpha_{s})\otimes x^{\lambda_{g}})\nonumber\\
&&+G(x,Q^{2})\frac{\partial}{\partial
\ln{x}}(C_{L,g}(x,\alpha_{s})\otimes x^{\lambda_{g}})\},\nonumber\\
\end{eqnarray}
where
$C_{L,g}(x,\alpha_{s})=\frac{\alpha_{s}}{4\pi}C^{LO}_{L,g}(x)+(\frac{\alpha_{s}}{4\pi})^{2}
C^{NLO}_{L,g}(x)+(\frac{\alpha_{s}}{4\pi})^{3}C^{NNLO}_{L,g}(x)$.
Here we used the Regge-like behavior of the gluon distribution
function in Eq.(8). Using Eqs.(8) and (9) and simplifying
derivative of the longitudinal structure function, we get
\begin{eqnarray}
\frac{\partial \ln F^{g}_{L}(x,Q^{2})}{\partial \ln{x}}=
\frac{\partial \ln G(x,Q^{2})}{\partial \ln{x}}+\frac{\partial \ln
I_{g}(x,Q^{2})}{\partial \ln{x}},
\end{eqnarray}
where $I_{g}(x,Q^{2})=<e^{2}> C_{L,g}(x,\alpha_{s})\otimes
x^{\lambda_{g}}$. We note that exponents $\lambda_{g}$ and
$\lambda_{L}$ are given as the derivatives
\begin{eqnarray}
\lambda_{g}=\frac{\partial{\ln}G(x,Q^{2})}{\partial{\ln(1/x)}}\nonumber\\
\mathrm{and}~~~~~~~~~~~~~~~~~~~\nonumber\\
\lambda_{L}=\frac{\partial{\ln}F_{L}(x,Q^{2})}{\partial{\ln(1/x)}}.
\end{eqnarray}
Therefore, the longitudinal exponent with respect to the gluonic
term is defined as follows
\begin{eqnarray}
\lambda_{L}=\lambda_{g}+\frac{\partial \ln
I_{g}(x,Q^{2})}{\partial \ln{(1/x)}},
\end{eqnarray}
here
\begin{eqnarray}
I_{g}(x,Q^{2})&=&<e^{2}>\int_{x}^{1}\frac{dz}{z}[\frac{\alpha_{s}}{4\pi}C^{LO}_{L,g}(x)+(\frac{\alpha_{s}}{4\pi})^{2}
C^{NLO}_{L,g}(x)\nonumber\\
&&+(\frac{\alpha_{s}}{4\pi})^{3}C^{NNLO}_{L,g}(x)]z^{\lambda_{g}}.
\end{eqnarray}

\subsection{2.2. Singlet+Gluon terms:}
The standard collinear factorization formula for the longitudinal
structure function in terms of singlet and gluon structure
function at small-$x$ is given by
\begin{eqnarray}
F_{L}(x,Q^{2})&=&C_{L,q}(\alpha_{s},x)\otimes
F_{2}^{s}(x,Q^{2})\nonumber\\
&&+<e^{2}>C_{L,g}(\alpha_{s},x)\otimes G(x,Q^{2}).
\end{eqnarray}
Taking the derivative of Eq.(14) with respect to $\ln{x}$ for each
value of constant $Q^{2}$, we get
\begin{eqnarray}
\frac{\partial F_{L}(x,Q^{2})}{\partial \ln{x}}&=&\frac{\partial
F_{2}(x,Q^{2})}{\partial \ln{x}}I_{s}(x,Q^{2})
+F_{2}^{s}(x,Q^{2})\frac{\partial I_{s}(x,Q^{2})}{\partial
\ln{x}}\nonumber\\
&&+\frac{\partial G(x,Q^{2})}{\partial \ln{x}}I_{g}(x,Q^{2})
G(x,Q^{2})\frac{\partial I_{g}(x,Q^{2})}{\partial
\ln{x}},\nonumber\\
\end{eqnarray}
where $I_{s}(x,Q^{2})=C_{L,q}(x,\alpha_{s})\otimes
x^{\lambda_{s}}$. Exploiting the small-$x$ behavior of the
distribution functions according to the hard-pomeron. Then
equation (14) can be rewritten as
\begin{eqnarray}
F_{L}(x,Q^{2})=F_{2}^{s}(x,Q^{2})I_{s}(x,Q^{2})+
G(x,Q^{2})I_{g}(x,Q^{2}).\nonumber\\
\end{eqnarray}
Now, using Eqs.(15) and (16), the longitudinal exponent
$\lambda_{L}$ is found directly from the singlet and gluon
exponents, namely
\begin{widetext}
\begin{eqnarray}
\lambda_{L}=\frac{I_{s}(x,Q^{2})\lambda_{s}+{\partial
I_{s}(x,Q^{2}) }/{\partial
\ln(1/x)}+K(x,Q^{2})[I_{g}(x,Q^{2})\lambda_{g}+{\partial
I_{g}(x,Q^{2}) }/{\partial \ln
(1/x)}]}{I_{s}(x,Q^{2})+K(x,Q^{2})I_{g}(x,Q^{2})},
\end{eqnarray}
\end{widetext}
where $K(x,Q^{2})=G(x,Q^{2})/F_{2}^{s}(x,Q^{2})$ [17]. We observe
that equation (17) implies a relationship between the longitudinal
exponent and singlet-gluon exponents for even $n_{f}$. Thus  an
analytical expression for the longitudinal
exponent $\lambda_{L}$ is suggested at LO, NLO and NNLO.\\
\subsection{2.3. Light+Charm terms:}
In a similar manner, the charm contribution to the longitudinal
structure function is considered and the longitudinal exponent can
be determined at small $x$ with help of the light and gluon
exponents as
\begin{eqnarray}
F^{\mathrm{Total}}_{L}=F^{\mathrm{Light}}_{L}(=F^{q}_{L}+F^{g}_{L})+F^{\mathrm{Heavy}}_{L}.
\end{eqnarray}
where $F^{\mathrm{Light}}_{L}$=Eq.(14) with $<e^{2}>=\frac{2}{9}$
for $n_{f}=3$ (number of active light flavours).\\
With respect to the recent measurements of HERA [18], the charm
contribution to the  structure function at small $x$ is a large
fraction of the total. This behavior is directly related to the
growth of the gluon distribution at small $x$ [11] as
\begin{eqnarray}
F_{L}^{c}(x,Q^{2},m^{2}_{c})&=&2e_{c}^{2}\frac{\alpha_{s}(\mu^{2})}{2\pi}\int_{1-\frac{1}{a}}^{1-x}dzC_{g,L}^{c}
(1-z,\zeta)\nonumber\\
&& {\times}G(\frac{x}{1-z},\mu^{2}),
\end{eqnarray}
where $a=1+4\zeta(\zeta{\equiv}\frac{m_{c}^{2}}{Q^{2}})$ and $\mu$
is the mass factorization scale. The factorization scale is equal
to the renormalization scales $\mu^{2}=4m_{c}^{2}$ or
$\mu^{2}=4m_{c}^{2}+Q^{2}$. Here $C^{c}_{g,L}$ is the charm
coefficient functions in LO and NLO analysis [19-21] as
\begin{eqnarray}
C_{g,L}(z,\zeta)&{\rightarrow}&C^{0}_{g,L}(z,\zeta)+a_{s}(\mu^{2})[C_{g,L}^{1}(z,\zeta)\\\nonumber
&&+\overline{C}_{g,L}^{1}(z,\zeta)ln\frac{\mu^{2}}{m_{c}^{2}}],
\end{eqnarray}
where $a_{s}(\mu^{2})=\frac{\alpha_{s}(\mu^{2})}{4\pi}$ and in the
NLO analysis
\begin{eqnarray}
\alpha_{s}(\mu^{2})=\frac{4{\pi}}{\beta_{0}ln(\mu^{2}/\Lambda^{2})}
-\frac{4\pi\beta_{1}}{\beta_{0}^{3}}\frac{lnln(\mu^{2}/\Lambda^{2})}{ln(\mu^{2}/\Lambda^{2})}
\end{eqnarray}
with $\beta_{0}=11-\frac{2}{3}n_{f}$ and
$\beta_{1}=102-\frac{38}{3}n_{f} $.\\
After doing the integration over $z$, Eq.(19) can be rewritten as
\begin{eqnarray}
F_{L}^{c}(x,Q^{2},m^{2}_{c})&=&G(x,Q^{2})[C_{g,L}^{c}(x,Q^{2})
{\otimes}x^{\lambda_{g}}]\nonumber\\
&&=G(x,\mu^{2})I_{c}(x,Q^{2}),
\end{eqnarray}
where
\begin{eqnarray}
I_{c}(x,Q^{2})=2e_{c}^{2}\frac{\alpha_{s}(\mu^{2})}{2\pi}\int_{1-\frac{1}{a}}^{1-x}C_{g,k}^{c}
(1-z,\zeta)(1-z)^{\lambda_{g}}dz.\nonumber\\
\end{eqnarray}
The $x$-derivative of the longitudinal structure function is
defined by
\begin{eqnarray}
\frac{\partial F_{L}(x,Q^{2})}{\partial \ln{x}}&=&Eq.(15)+
\frac{\partial F_{L}^{c}(x,Q^{2})}{\partial \ln{x}}=Eq.(15)+\nonumber\\
&&\frac{\partial G(x,Q^{2})}{\partial \ln{x}}I_{c}(x,Q^{2})
+G(x,Q^{2})\frac{\partial I_{c}(x,Q^{2})}{\partial
\ln{x}}.\nonumber\\
\end{eqnarray}
Following the suggestion of the power-like behavior of the
logarithmic $x$-derivative of the distribution functions we have
the longitudinal exponent $\lambda_{L}$ for
$n_{f}=3+\mathrm{charm}$, as
\begin{widetext}
\begin{eqnarray}
\lambda_{L}=\frac{I_{s}(x,Q^{2})\lambda_{s}+{\partial
I_{s}(x,Q^{2}) }/{\partial
\ln(1/x)}+K(x,Q^{2})[(I_{g}(x,Q^{2})+I_{c}(x,Q^{2})
)\lambda_{g}+{\partial}/{\partial \ln
(1/x)}(I_{g}(x,Q^{2})+I_{c}(x,Q^{2}))]}{I_{s}(x,Q^{2})+K(x,Q^{2})[I_{g}(x,Q^{2})+I_{c}(x,Q^{2})]}.\nonumber\\
\end{eqnarray}
\end{widetext}
Therefore, equations (12), (17) and (25) are a set of formulas to
extract the longitudinal exponent from the singlet and gluon
exponents at gluonic, singlet+gluon and light +charm terms
respectively in LO, NLO and NNLO.\\
We now discuss how the presented results give the exponents for
the longitudinal structure functions at small $x$. In Ref.[4] the
authors have suggested that singlet and gluon effective exponents
can be reasonably defined by color dipole model and hard-pomeron
exponents. The exponents of $\lambda_{s}$ and $\lambda_{g}$ are
found to be $\simeq 0.33$ and $\simeq 0.42$ respectively [4].
Based on the coefficient functions [9] and effective exponents we
present result for the longitudinal exponents at LO, NLO and NNLO
using
Eqs.(12), (17) and (25) respectively.\\
In Eq.(12) the longitudinal exponent behavior for the gluonic
contribution is determined. After doing the integration and using
the required coefficient functions the longitudinal exponents, in
the range $10^{-5} {\leq} x {\leq} 10^{-2}$ and $2 {\leq} Q^{2}
{\leq}45~ GeV^{2}$, are determined in Fig.1. In this figure the
obtained results are compared with $\lambda_{g}=0.42$. We observe
that $\lambda_{L}\leq \lambda_{g}$ at NLO and NNLO analysis. In
all the graphs, $\lambda_{L}$  is equal to $\lambda_{g}$ at very
low $x$ values. For all values of $x$ we observe that
$\lambda_{L}= \lambda_{g}$ only at LO analysis. In this case the
longitudinal exponent is hard-pomeron dominated. Therefore the
averaged value to all exponents has the effective constraint where
 the  effective longitudinal exponent has the following value as
\begin{eqnarray}
\lambda_{L}^{Gluonic}\simeq 0.41.
\end{eqnarray}
In the following the longitudinal exponent is obtained using the
singlet and gluon terms from Eq.(17) with respect to the exponents
in Fig.2. In this figure, the longitudinal exponent $\lambda_{L}$
is plotted against $x$ for different values of $Q^{2}$ in
comparison with singlet  ($\lambda_{s}=0.33 $) and gluon
($\lambda_{g}=0.42 $) exponents at LO, NLO and NNLO analysis.
Since $\lambda_{L}$ is an analytical function of $x$, it can not
be exactly constant at small $x$. This is due to the coefficient
function  and dispersion of data. Nevertheless, we observe that
$\lambda_{L}$ does not strongly depends on $x$ at $x<0.01$ [22].
In fact, it is more likely that exponent depends weakly on $x$.
However the averaged value to all exponents  in this case has the
following value at NNLO as
\begin{eqnarray}
\lambda_{L}^{Singlet+Gluon}\simeq 0.40.
\end{eqnarray}
In Fig.3, the values of longitudinal exponent are shown as a
function of $x$ at four different fixed $Q^{2}$ values with
respect to the light ($n_{f}=3$)+charm coefficient functions.
After doing some derivation of the heavy quarks  we observe that
the longitudinal exponents have the same behavior as discussed in
Figs.1 and 2. The merit of this plot, in comparison with another
one, is mainly due to its relation with the charm distribution.
The data have the property that the charm structure function
require a hard-pomeron component [1-2,5,18]. The averaged value to
all exponents  for the light and charm distribution at NNLO has
the following value
\begin{eqnarray}
\lambda_{L}^{Light+charm}\simeq 0.38.
\end{eqnarray}
It can be clearly seen that the longitudinal exponents decrease as
active flavours increases, but with a somewhat smaller rate. It
can be well described by
\begin{eqnarray}
\lambda_{s}{<}\lambda_{L}^{Light+charm}<
\lambda_{L}^{Singlet+Gluon} < \lambda_{L}^{Gluonic}
{\leq}\lambda_{g}.
\end{eqnarray}
Furthermore, these solutions predict that $\lambda_{L}\neq
\lambda_{g}$ in a wide range of $x-Q^{2}$ values at high order
corrections.\\
In Fig.4 we compare these predictions for longitudinal exponents
as a function of $Q^{2}$. The exponent $\lambda_{L}$ of the
longitudinal structure function is observed that depends weakly on
$Q^{2}$. It can be represented by a constant $\lambda_{L}$ which
is almost independent of $x$ and $Q^{2}$. This is consistent with
the hard-pomeron defined by  Donnachie and Landshoff [1-2,5]. So
the simplest form to the small $x$ behavior of the longitudinal
structure function corresponds to $F_{L}\sim x^{-\lambda_{L}}$.
Having conclude that the data for $F_{L}$ require a hard pomeron
component with condition of Eq.(29).\\

\subsection{3. Reduced cross section}
The extraction of the reduced double differential cross section is
based  on two proton structure functions $F_{2}(x,Q^{2})$ and
$F_{L}(x,Q^{2})$. When $y\rightarrow 1$, the reduced cross section
$\sigma_{r}$ tends to $F_{2}-F_{L}$. An important advantage of
HERA is used to perform an extraction of the longitudinal
structure function with respect to the extrapolation
 and  derivative methods [15,23].\\
 As discussed in section 2, the behavior of the proton
 structure functions $F_{2}$ and $F_{L}$ are
 $x^{-\lambda_{s}}$ and $x^{-\lambda_{L}}$ at fixed $Q^{2}$
 respectively. On this basis the reduced cross section
 distribution can be parametrised as
\begin{eqnarray}
\sigma_{r}=
A_{s}(Q^{2})x^{-\lambda_{s}}-\frac{y^{2}}{Y_{+}}A_{L}(Q^{2})x^{-\lambda_{L}}.
\end{eqnarray}
We analysis the reduced cross section behavior with a power-like
behavior at small $x$ at fixed $Q^{2}$ as
\begin{eqnarray}
\sigma_{r}(x,Q^{2})\equiv A_{\sigma}(Q^{2})x^{-\lambda_{\sigma}},
\end{eqnarray}
where
\begin{eqnarray}
\lambda_{\sigma}=\frac{\partial{\ln}\sigma_{r}(x,Q^{2})}{\partial{\ln(1/x)}}.
\end{eqnarray}
In order to do this, the derivative of $\ln{\sigma_{r}}$, taken at
fixed $Q^{2}$, is given by
\begin{widetext}
\begin{eqnarray}
\frac{\partial{\ln}\sigma_{r}(x,Q^{2})}{\partial{\ln
x}}|_{Q^{2}}=[ {\frac{\partial{\ln}F_{2}(x,Q^{2})}{\partial{\ln
x}}-\frac{F_{L}}{F_{2}}\frac{y^{2}}{Y_{+}}\frac{\partial{\ln}F_{L}(x,Q^{2})}{\partial{\ln
x}}-\frac{F_{L}}{F_{2}}\frac{\partial}{\partial{\ln
x}}\frac{y^{2}}{Y_{+}}}]/[{1-\frac{F_{L}}{F_{2}}\frac{y^{2}}{Y_{+}}}].
\end{eqnarray}
\end{widetext}
Hence, the reduced cross section exponent is defined by an
analytical expression as
\begin{eqnarray}
\lambda_{\sigma}=[\lambda_{s}-\lambda_{L}
\frac{F_{L}}{F_{2}}\frac{y^{2}}{Y_{+}}+\frac{F_{L}}{F_{2}}\frac{\partial}{\partial{\ln
x}}\frac{y^{2}}{Y_{+}}]/[{1-\frac{F_{L}}{F_{2}}\frac{y^{2}}{Y_{+}}}],
\end{eqnarray}
when $0<y<1$.
 For $y{\rightarrow}0$ the reduced cross section
exponent tends to the limit
\begin{eqnarray}
\lambda_{\sigma}{\rightarrow}\lambda_{s},
\end{eqnarray}
and tends to the limit
\begin{eqnarray}
\lambda_{\sigma} \simeq \frac{\lambda_{s}-\lambda_{L}
\frac{F_{L}}{F_{2}}}{1-\frac{F_{L}}{F_{2}}},
\end{eqnarray}
when $y{\rightarrow}1$. We note that the behavior of
$\frac{\partial}{\partial{\ln x}}\frac{y^{2}}{Y_{+}}$ in Eq.(34)
is controlled at two limited region (Eqs.(35) and (36)). In Fig.5,
 the behavior of $\frac{\partial}{\partial{\ln
x}}\frac{y^{2}}{Y_{+}}$ at fixed $s$ and $Q^{2}$ values is shown
that lead to rapid depletion and enhancement in the small-$x$
region ($10^{-6}<x<10^{-3}$). To better illustration this behavior
at small $x$, the reduced cross section exponent
$\lambda_{\sigma}$ is plotted versus the $x$ variable (see Fig.6).
It can be clearly seen that this result is dependence to the ratio
of the structure functions behavior. In color dipole model
[24-25], a strict bound for the ratio $F_{L}/F_{2}$ is defined as
$\frac{F_{L}}{F_{2}}\leq 0.27$. For realistic dipole-proton
cross-section [26] the bound is reduced to $0.22$. From the new
measurement of $F_{L}$ at HERA, a phenomenological model derive
ratio of structure functions where lead to the bound $0.12$ in a
wide range of $Q^{2}$ values [17]. In Fig.6 the effects of these
bounds for the reduced cross section exponent have been presented.
For a constant $Q^{2}$, the reduced cross section exponent has the
same behavior of the singlet exponent at $x>10^{-3}$ and
$x<10^{-5}$. There is some violation at $10^{-5}<x<10^{-3}$. In
this range  a depletion and then an enhancement is observable in
all figures as $x$ decreases.\\
In Fig.7 the form $x^{-\lambda_{\sigma}}\equiv
\frac{\sigma_{r}(x,Q^{2})}{A_{\sigma}(Q^{2})}$ for the reduced
cross section parametrization at small $x$ is plotted. For fixed
$Q^{2}$, the reduced cross section at HERA data [15] rises with
decreasing $x$ as $x\rightarrow 10^{-3}$. The increase of $F_{ L}$
towards small-$x$ is consistent with the high-order QCD
corrections. This behavior is reflecting the decrease of the
reduced cross section towards small-$x$. In Fig.7 this
characteristic of the reduced cross section is observed with
respect to the depletion behavior at this region. This behavior is
consistence with the available HERA data [15]. Thus we observe a
continuous increase then decrease
towards small $x$.\\
In H1 analysis, the measured reduced cross section is represented
as
\begin{eqnarray}
\sigma_{r}(x,Q^{2})\equiv
F_{2}(x,Q^{2})[1-\frac{y^{2}}{Y_{+}}\frac{R}{1+R}],
\end{eqnarray}
where the value $R$ is generally assumed that is constant for all
$Q^{2}$ bins. In HERA analysis, the observations obtained with the
general methods such as derivative method, offset method and
fitted method [15]. We now discuss how the presented results give
an analytical analysis for the ratio $F_{2}/\sigma$ with respect
to the effective exponents at small $x$. In order to obtain the
ratio $F_{2}/\sigma$, the derivative of the reduced cross section,
taken at fixed $Q^{2}$, is used as
\begin{eqnarray}
\frac{\partial \sigma_{r}}{\partial{\ln}y}|_{Q^{2}}=\frac{\partial
F_{2}}{\partial{\ln}y}|_{Q^{2}}-F_{L}.2y^{2}.\frac{2-y}{Y_{+}}-\frac{\partial
F_{L}}{\partial{\ln}y}.\frac{y^{2}}{Y_{+}}.
\end{eqnarray}
Using the fact that cross section and distribution functions have
a power-law behavior with an effective exponent. Considering the
relationship between the functions and effective exponents, shows
a similar relation as we have
\begin{eqnarray}
\lambda_{\sigma} \sigma_{r}=\lambda_{s}
F_{2}-F_{L}.2y^{2}.\frac{2-y}{Y_{+}}-\lambda_{L}
F_{L}.\frac{y^{2}}{Y_{+}}.
\end{eqnarray}
Now, using Eqs.(7) and (39), the ratio $F_{2}/\sigma$ is found
directly from the exponents, namely
\begin{eqnarray}
\frac{F_{2}}{\sigma}=\frac{\lambda_{\sigma}-[\lambda_{L}+\frac{2}{Y_{+}}(2-y)]}{\lambda_{s}-[\lambda_{L}+\frac{2}{Y_{+}}(2-y)]}.
\end{eqnarray}
Here we used the pomeron value of the exponents assumed for small
$x$ region by the available  H1 data. In Fig.8 a comparison is
made between our obtained values and the available data. The
results of analytical solutions with respect to the exponents for
the ratio $F_{2}/\sigma$ clearly show significant agreement over a
wide range of $x$ and $Q^{2}$ values. At very small $x$ the
nonlinear corrections
 have to be take into account. Extension of current result to the
 nonlinear effect is also a valuable task to follow it in
 future.\\

\subsection{4. Conclusion}
In this section, a set of new formulas connecting the longitudinal
exponent with the gluon and singlet exponents at small $x$ have
been presented. Based upon the hard pomeron behavior of the gluon
and singlet exponents, the behavior of the longitudinal exponent
at high order corrections is considered. We found that
longitudinal exponent behavior is dependence to the active
flavors. The value of the longitudinal exponent is similar to the
one predicted for the singlet and gluons. This exponent is almost
independent of $x$ for $x<10^{-2}$. We see that
$\lambda_{s}<\lambda_{L} \leq \lambda_{g}$. This exponent as a
function of $Q^{2}$ is consistent with the hard pomeron behavior.
Thus the behavior of $F_{L}$ at small $x$ is consistent with a
dependence $F_{L}(x,Q^{2})=A_{L}(Q^{2})x^{-\lambda_{L}}$
throughout that region.\\
Also we analyse the behavior of the exponent for  the reduced
cross section. The behavior of $\Delta(\equiv
\frac{\partial}{\partial \ln x}\frac{y^{2}}{Y_{+}})$ at high  and
very low-$x$ values is considered as this behavior is linear and
 equal to zero. But in the region $10^{-6}<x<10^{-3}$ (at four
$Q^{2}$ value determined), the behavior of this function
($\Delta$) can no longer be neglected. The deviation of this
expression from zero shows the importance of non-linear effects. A
depletion in the low $x$ (high $y$) is called shadowing whereas
an enhancement is called anti-shadowing [27].\\
The oscillating behavior for $\lambda_{\sigma}$ can be explained
by new effects at low-$x$, such as  the nonlinear recombination.
The behavior of the function $x^{-\lambda_{\sigma}}$ increase as
$x$ decreases. The negative shadowing and the positive
anti-shadowing corrections to this behavior can be explain by the
non- linear effects to the structure functions.  In view of these
results for the
 exponents, we may infer some evidence for non- linear effects at LHeC [28].\\
Considering these determined exponents and using the derivatives
methods to find the ratio $\frac{F_{2}}{\sigma}$ and finally
comparing with the H1 data, one concludes that this new method is
capable of determining the ratio $\frac{F_{2}}{\sigma}$ with
considerable precision.\\

\subsection{References}

1. A.Donnachie and P.V.Landshoff, Phys.Lett.B {\bf437}, 408(1998 ).\\
2. J.R.Cudell, A.Donnachie and P.V.Landshoff, Phys.Lett.B {\bf448}, 281(1999).\\
3. K Golec-Biernat and A.M.Stasto, Phys.Rev.D {\bf80},
014006(2009).\\
4. B.Rezaei and G.R.Boroun, arXiv[hep-ph]:1811.02785(2018).\\
5. A.Donnachie and P.V.Landshoff, Phys.Lett.B {\bf550}, 160(2002 )\\; P.V.Landshoff,hep-ph/0203084; D.Britzger et al., arXiv:1901.08524 (2019).\\
6.M.Hentschinski et al., Phys.Rev.Lett.{\bf110}, 041601(2013);
B.Rezaei and G.R.Boroun, Int.J.Theor.Phys.{\bf57}, 2309(2018).\\
7. G.R.Boroun, Phys.Rev.C{\bf97}, 015206(2018);
Eur.Phys.J.Plus{\bf129}, 19(2014); G.R.Boroun, B.Rezaei and
J.K.Sarma, Int.J.Mod.Phys.A{\bf29}, 1450189(2014); N.Baruah et al., Int.J.Theor.Phys.{\bf54}, 3596(2015).\\
8. G.Altarelli and G. Martinelli, Phys.Lett.B{\bf 76}, 89(1978).\\
9. S.Moch and J.A.M.Vermaseren, A.Vogt, Phys.Lett.B{\bf 606}, 123(2005); S.Alekhin et al., arXiv:1808.08404.\\
10. E. Laenen et al., Nucl.Phys.B{\bf 392}, 162(1993); S.Riemersma
et al., Phys.Lett.B{\bf 347}, 143(1995); A.V.Kisselev, Phys.Rev.D{60}, 074001(1999).\\
11. G.R.Boroun and B.Rezaei, Nucl.Phys.B{\bf857}, 143(2012);
EPL{\bf100}, 41001(2012); G.R.Boroun, Nucl.Phys.B{\bf884}, 684(2014); N.Ya.Ivanov, Nucl.Phys.B{\bf814}, 142(2009).\\
12. R.S.Thorne, arXiv:hep-ph/9805298(1998); L.A.Harland-Lang et al., Eur.Phys.J.C{\bf76}, 10(2016).\\
13.  Yu.L.Dokshitzer, Sov.Phys.JETP {\textbf{46}}, 641(1977);
G.Altarelli and G.Parisi, Nucl.Phys.B \textbf{126}, 298(1977);
V.N.Gribov and L.N.Lipatov, Sov.J.Nucl.Phys. \textbf{15},
438(1972).\\
14. N.Ghahramany and G.R.Boroun, Phys.Lett.B{\bf 528}, 239(2002);
G.R.Boroun, Lith.J.Phys.{\bf48}, 121(2008).\\
15. H1 Collab. (C.Adloff et al.), Eur.Phys.J.C{\bf21}, 33(2001);
ZEUS Collab. (S.Chekanov et al.), Phys.Lett.B{\bf682}, 8(2009);
H1 and ZEUS Collab. (H.Abromowicz et al.), Eur.Phys.J.C{\bf75}, 580(2015).\\
16.  G.R.Boroun and B.Rezaei, Eur.Phys.J.C{\bf72}, 2221(2012); R.Fiore et al., JETP Lett.{\bf90}, 319(2009); J.Sheibani et al., Phys.
Rev.C{\bf98}, 045211(2018).\\
17. G.R.Boroun and B.Rezaei, arXiv:1901.05199(2019); L.P.Kaptari et al., arXiv:1812.00361 (2018).\\
18. H1 Collab. (F.D.Aaron et al.), Eur.Phys.J.C{\bf65}, 89(2010).\\
19. M.Gluk, E.Reya and A.Vogt, Z.Phys.C\textbf{67}, 433(1995); Eur.Phys.J.C\textbf{5}, 461(1998).\\
20. E.Laenen, S.Riemersma, J.Smith and W.L. van Neerven,
Nucl.Phys.B \textbf{392}, 162(1993); A.~Y.~Illarionov,
B.~A.~Kniehl and A.~V.~Kotikov, Phys.\
Lett.\  B {\bf 663}, 66 (2008).\\
21. S. Catani, M. Ciafaloni and F. Hautmann, Preprint
CERN-Th.6398/92, in Proceeding of the Workshop on Physics at HERA
(Hamburg, 1991), Vol. 2., p. 690; S. Catani and F. Hautmann, Nucl.
Phys. B \textbf{427}, 475(1994); S. Riemersma, J. Smith and W. L.
van Neerven, Phys. Lett. B \textbf{347}, 143(1995).\\
22. P.Desgrolard et al., JHEP{\bf02}, 029(2002); A.A.Godizov, Nucl.Phys.A{\bf927}, 36(2014);
 A.Watanabe and K.Suzuki, Phys.Rev.D{\bf86}, 035011(2012).\\
23. H1 Collab. (V.Andreev et al.), Eur.Phys.J.C{\bf74},
2814(2014); E.M.Lobodzinska, arXiv:hep-ph/0311180(2003); V.Tvaskis et al., Phys.Rev.C{\bf97}, 045204(2018);
H1 Collab. (F.D.Aaron et al.), Eur.Phys.J.C{\bf71}, 1579(2011); P.Monaghan et al., Phys.Rev.Lett.{\bf110}, 152002(2013).\\
24. C.Ewerz et al., Phys.Lett.B{\bf720}, 181(2013); M.Kuroda and D.Schildknecht, Phys.Rev.D{\bf85}, 094001(2012).\\
25. C.Ewerz et al., JHEP{\bf1103}, 062(2011).\\
26. M.Niedziela and M.Praszalowicz, Acta Phys.Polon.B{\bf46},
2019(2015).\\
27. W.Zhu, et.al., Nucl.Phys.B {\bf551}, 245(1999);
Nucl.Phys.B{\bf559}, 375(1999); J.Ruan, et.al.,
Nucl.Phys.B{\bf760}, 128(2007).\\
28. You Yu et al., J.Phys.G{\bf45}, 125003(2018); Hao Sun Pos DIS2017(2018)104; Pos DIS2018(2018)186; Lin Han et al., Phys.Lett.B{\bf771}, 106(2017);
 Yao-Bei Liu, Nucl.Phys.B{\bf923}, 312(2017); Yan-Ju Zhang et al., Phys.Lett.B{\bf768}, 241(2017); M.Mangano (CERN) et al., CERN-ACC-2018-0056;
 M.Klein, Annalen Phys.{\bf528}, 138(2016).\\

\begin{figure} \centering
\includegraphics[width=0.5\textwidth]{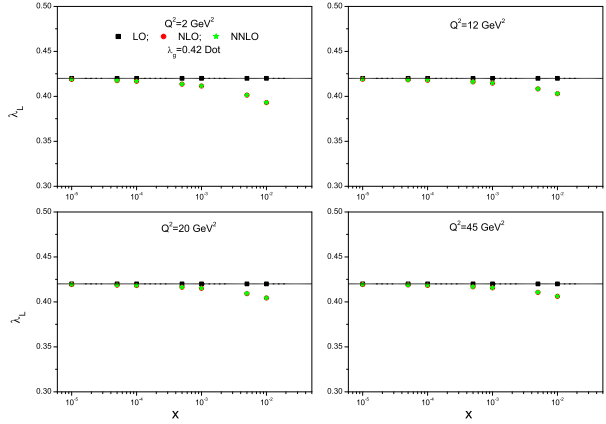}
\caption{The longitudinal exponent $\lambda_{L}$ given by Eq.(12)
versus $x$ at four fixed $Q^{2}$ values at LO, NLO and NNLO
analysis, compared with the gluon exponent $\lambda_{g}=0.42$ (Dot
line).} \label{Fig1}
\end{figure}
\begin{figure} \centering
\includegraphics[width=0.5\textwidth]{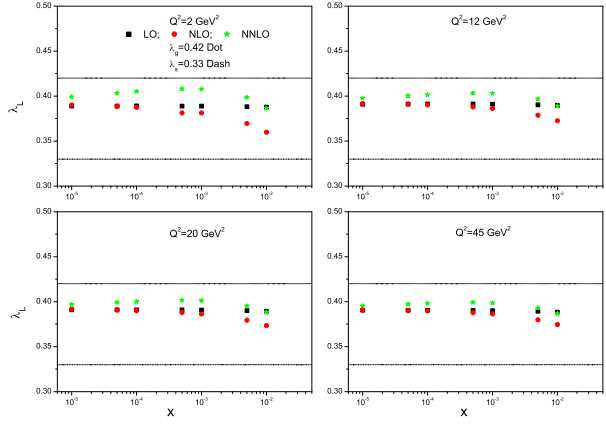}
\caption{The longitudinal exponent $\lambda_{L}$ given by Eq.(17)
versus $x$ at four fixed $Q^{2}$ values at LO, NLO and NNLO
analysis, compared with the gluon exponent $\lambda_{g}=0.42$ (Dot
line) and singlet exponent $\lambda_{s}=0.33$ (Dash line).}
\label{Fig1}
\end{figure}
\begin{figure} \centering
\includegraphics[width=0.5\textwidth]{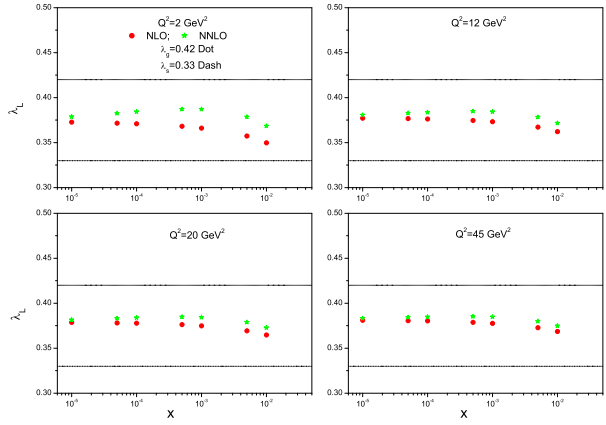}
\caption{The longitudinal exponent $\lambda_{L}$ given by Eq.(25)
versus $x$ at four fixed $Q^{2}$ values at NLO and NNLO analysis,
compared with the gluon exponent $\lambda_{g}=0.42$ (Dot line) and
singlet exponent $\lambda_{s}=0.33$ (Dash line). The
renormalization scale is $\mu=\sqrt{4m_{c}^{2}+Q^{2}}$. }
\label{Fig1}
\end{figure}
\begin{figure} \centering
\includegraphics[width=0.5\textwidth]{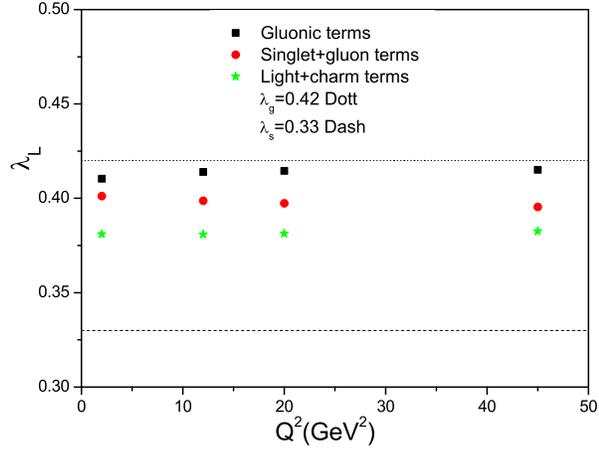}
\caption{The longitudinal exponent $\lambda_{L}$ plotted against
$Q^{2}$ with respect to the gluonic terms, singlet+gluon terms and
light+charm terms, compared with the gluon exponent
$\lambda_{g}=0.42$ (Dot line) and singlet exponent
$\lambda_{s}=0.33$ (Dash line).} \label{Fig1}
\end{figure}
\begin{figure} \centering
\includegraphics[width=0.5\textwidth]{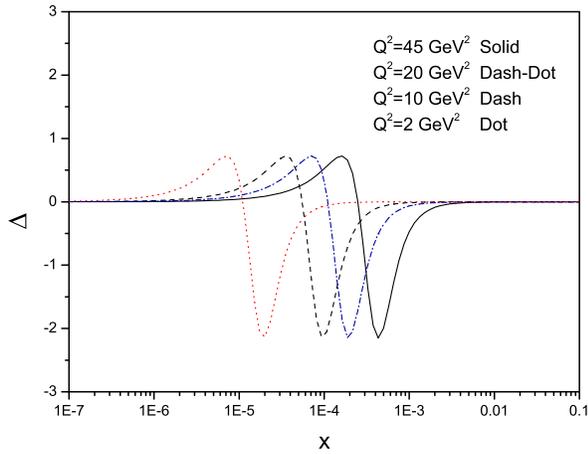}
\caption{The calculated values
$\Delta=\frac{\partial}{\partial{\ln x}}\frac{y^{2}}{Y_{+}}$
plotted against $x$ at four fixed $Q^{2}$ values at
$\sqrt{s}=300.9~ GeV$ (H1 2001 [15]).} \label{Fig1}
\end{figure}
\begin{figure} \centering
\includegraphics[width=0.5\textwidth]{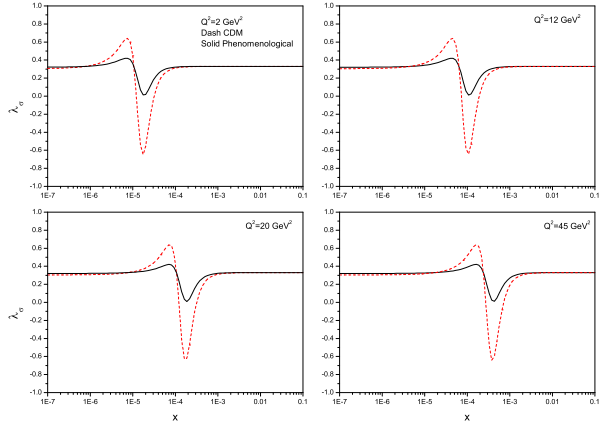}
\caption{The behavior of the reduced cross section exponent
$\lambda_{\sigma}$  plotted against $x$ at four fixed $Q^{2}$
values with respect to the ratio of the structure functions
$F_{L}/F_{2}$ in color dipole model (Dash line) [24] and
phenomenological model (Solid line) [17].} \label{Fig1}
\end{figure}
\begin{figure} \centering
\includegraphics[width=0.5\textwidth]{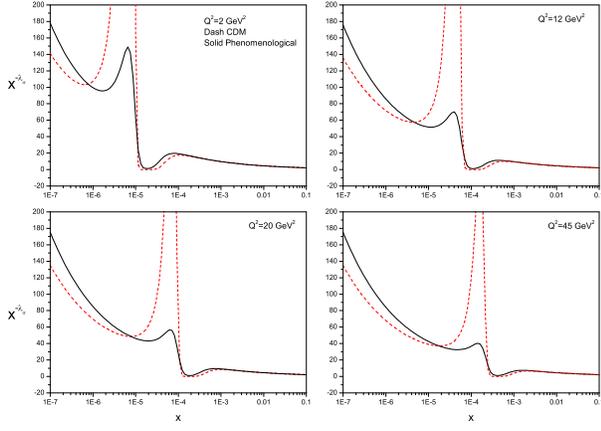}
\caption{The function
$x^{-\lambda_{\sigma}}\equiv\frac{\sigma_{r}(x,Q^{2})}{A_{\sigma}(Q^{2})}$
as  a function of $x$ for different $Q^{2}$ bins at the same
models for the ratio of structure functions in Fig.6.}
\label{Fig1}
\end{figure}
\begin{figure} \centering
\includegraphics[width=0.5\textwidth]{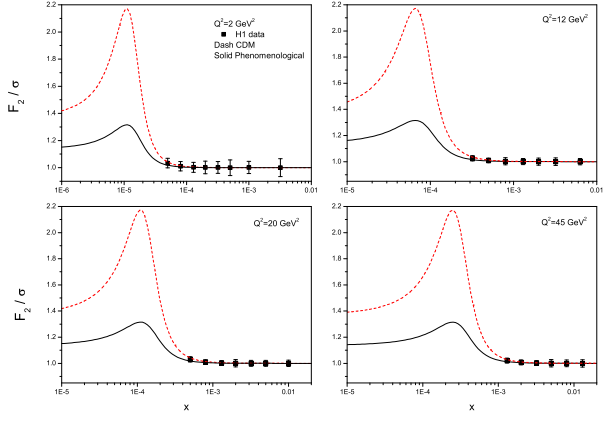}
\caption{Our results for the ratio $\frac{F_{2}}{\sigma}$, using
Eq.(40) and its comparison with  the H1 2001 data [15] as
accompanied with total errors.} \label{Fig1}
\end{figure}
\end{document}